\documentstyle[aps,preprint,tighten,epsfig]{revtex}
\def\beq{\begin{equation}}
\def\eeq{\end{equation}}
\def\beqa{\begin{eqnarray}}
\def\eeqa{\end{eqnarray}}

\def\GeV{\nobreak\,\mbox{GeV}}

\def\maior{\smash{\mathop{>}\limits_{\raise4pt\hbox{$\scriptstyle \sim$}}}}
\def\menor{\smash{\mathop{<}\limits_{\raise4pt\hbox{$\scriptstyle \sim$}}}}

\begin{document}
%\draft 
%%%%%%%%% %%%%%%%%% %%%%%%%%% %%%%%%%%% %%%%%%%%% %%%%%%%%% %%
\title{ \bf {Chiral Symmetry in Charmonium - Pion Cross Section}}
\author{F.S. Navarra\thanks{e-mail: navarra@if.usp.br},
M. Nielsen\thanks{e-mail: mnielsen@if.usp.br} and 
M.R. Robilotta\thanks{e-mail: mane@if.usp.br}\\
{\it Instituto de F\'{\i}sica, Universidade de S\~{a}o Paulo}\\
{\it C.P. 66318,  05315-970 S\~{a}o Paulo, SP, Brazil}} 
\maketitle
\vspace{1cm}
\begin{abstract}
We perform a non-perturbative calculation of the $J/\Psi-\pi$ cross 
section using a $SU(2)\times SU(2)$ effective Lagrangian. Our results differ
from those of previous calculations, specially in the description of
vertices involving pions.
\noindent

PACS: 12.39.Fe~~13.85.Fb~~14.40.Lb
\end{abstract}

\vspace{1cm}
%\newpage

%\section{Introduction}

Reliable values for the 
charmonium -  hadron cross sections are of crucial importance in the context 
of quark- gluon plasma  physics. Part of these interactions happens in the 
early stages of
the nucleus- nucleus collisions and therefore  at high energies 
($\sqrt{s} \simeq 10 - 20$ GeV) and one may try to apply perturbative QCD. 
On the other hand, a significant part of the charmonium - hadron interactions 
occurs
when other light particles have already been produced, forming a ``fireball''. 
Interactions inside this fireball happen at much lower energies 
($\sqrt{s} \le 5$ GeV) and one has to apply non-perturbative methods.

One possible reaction mechanism is meson exchange, that can be studied by
means of effective Lagrangians, constrained by flavor and
chiral symmetries as well as gauge invariance. This approach has been 
pioneered by  Matinyan and B. M\"uller \cite{mamu98} and further developed by 
three other groups \cite{osl,haglin,haga,lkz,linko,ldk}. 

In the first work, only $D$ meson exchange was considered and the authors
obtained a small cross section of order of $2.5$ mb at $\sqrt{s}=5 \GeV$
for the process $\pi+J/\psi\rightarrow D+D^*$. In Refs. 
\cite{haglin,haga} $D^*$ meson exchange and four-point couplings were 
included, the latter in order to preserve gauge invariance. This led to
a much larger cross section, of around $30$ mb at $\sqrt{s}=5 \GeV$, 
mainly
due to the inclusion of the $D^*$ meson exchange. The contribution of the
four-point diagram to the cross section is of the same order as the process
with the $D$ meson exchange. In Ref.~\cite{osl} anomalous parity interactions
were included. In particular, anomalous term such as $D^*D^*\pi$ opens new
absorption channels not included before. In the case of the channel
$\pi+J/\psi\rightarrow D+D^*$, the inclusion of the $D^*D^*\pi$ vertex 
increases even more the cross section, that becomes about
$100$ mb at $\sqrt{s}=5 \GeV$. The inclusion of new absorption channels 
was also the idea guiding the work presented in Ref.~\cite{haga}. 
As an example,  the anomalous process $\pi+J/\psi
\rightarrow \eta_c\rho$ alone gives a cross section of order
of $60$ mb at $\sqrt{s}=5 \GeV$. 

A common feature of all these works is that they
take $SU(4)\times SU(4)$ symmetry as the point of
departure for describing mesonic interactions. As this symmetry is badly
broken in nature, at some stage of the calculation one is forced to use
the empirical values of the relevant masses. In this work we show that this
procedure may lead to inconsistencies which are numerically important.

%\section{Lagrangians and equations of motion}

Let us consider, for instance, the diagram for the process 
$\pi+J/\psi\rightarrow D+D^*$ with the $D^*$ meson exchange, given in Fig.~1.
It is based on a vector interaction among three spin one mesons, and the axial
 coupling between two pseudoscalar mesons and the $D^*$. For the former, one
uses the Lagrangian of Refs.~\cite{osl,linko}, given by
\begin{eqnarray}
{\cal L}_{\psi D^*D^*}&=& ig_{\psi D^*D^*}~
\left [ \psi^\mu \left ( \partial_\mu D^{* \nu} \bar {D^*_\nu} 
- D^{* \nu} \partial_\mu \bar {D^*_\nu} \right )
+\left ( \partial_\mu \psi^\nu D^*_\nu -\psi^\nu \partial_\mu D^*_\nu \right )
\bar {D^{* \mu}} \right .\nonumber \\ 
&+&\left . D^{* \mu} \left ( \psi^\nu \partial_\mu \bar {D^*_\nu} -
\partial_\mu \psi^\nu \bar {D^*_\nu} \right ) \right ] ~ . 
\label{jdsds} 
\end{eqnarray}

For the axial vertex one finds two alternative forms in the literature,
namely \cite{mamu98,linko}
\begin{eqnarray}
{\cal L}_{\pi DD^*}^{(I)}&=&{i\over2}g_{\pi DD^*}~ \left[\left(
\bar D  \vec \tau D^{* \mu} - \bar{D^{* \mu}} \vec \tau D\right)
\cdot \partial_\mu \vec \pi  
-\left(\partial_\mu \bar{D}\vec \tau D^{* \mu} -  \bar{D^{* \mu}} 
\vec \tau\partial_\mu D\right) \cdot  \vec \pi\right]~,
\label{jddko} 
\end{eqnarray}
and  \cite{osl}
\begin{eqnarray}
{\cal L}_{\pi DD^*}^{(II)}&=&ig_{\pi DD^*}~ \left(
\bar D  \vec \tau D^{* \mu} - \bar{D^{* \mu}} \vec \tau D\right)
\cdot \partial_\mu \vec \pi ~.
\label{jddsu} 
\end{eqnarray}

In the above equations, $\vec \tau$ are the Pauli matrices, 
and $\vec \pi$ denote the pion  meson isospin triplet, 
while $D\equiv (D^0,D^+)$ and $D^*\equiv (D^{*0},D^{*+})$ 
denote the pseudoscalar and vector charm meson doublets, respectively.
The Lagrangians in Eqs.~(\ref{jddko}) and (\ref{jddsu}) may be related by 
performing an integration by parts into the last term in Eq.(\ref{jddko}),
which allows us to write
\beq
{\cal L}_{\pi DD^*}^{(I)}={\cal L}_{\pi DD^*}^{(II)}+{i\over2}g_{\pi DD^*}~ 
\left(\bar D  \vec \tau \partial_\mu D^{* \mu}-\partial_\mu \bar{D^{* \mu}} 
\vec \tau D\right) \cdot  \vec \pi ~.
\label{rel}
\eeq

This result indicates that the two forms of the $\pi DD^*$ interaction would 
be equivalent if the condition $\partial_\mu D^{* \mu}=
\partial_\mu \bar{D^{* \mu}}=0$ hold, which is the case for on mass shell
vector mesons. However, in the sequence, we show that this condition is
not valid in the presence of interactions, and
hence that Eqs.~(\ref{jddko}) and (\ref{jddsu}) correspond to different 
dynamical hypotheses.

We begin by considering the specific case of the process displayed in Fig.~1.
Calling the four-momenta of the initial mesons $\pi$ and  $J/\psi$ by $p_1$ 
and $p_2$ and those of the final mesons $D^*$ and 
$D$ by $p_3$ and $p_4$ respectively, the vector vertex between the three
vector mesons is given by
\beq
i\Gamma^\beta=ig_{\psi D^*D^*}\epsilon_{2\mu}\epsilon_{3\nu}^*~\left(
-2p_3^\mu~ g^{\beta\nu}-2p_2^\nu~ g^{\beta\mu}+(p_2+p_3)^\beta~ 
g^{\mu\nu}\right)~,
\label{veve}
\eeq
where $\epsilon_i$ is the polarization vector of the vector meson with 
momentum $p_i$, and we have already used the orthogonality relation for 
vector mesons:
$\epsilon_2^\mu~p_{2\mu}~=~\epsilon_3^\nu~p_{3\nu}~=~0$. For the remaining
part of the diagram, which includes the $D^*$ propagator $(\Delta_{\alpha\beta}
)$ and the axial vertex ($A^\alpha$), we get, using Eq.~(\ref{jddko}):
\beqa
iA_\alpha^{(I)}~i\Delta^{\alpha\beta} &=&{i\over2}g_{\pi DD^*}~ (p_1+
p_4)_\alpha~i\Delta^{\alpha\beta}
\nonumber\\
%&=&-g_{\pi DD^*}~ \left(p_1-{q\over2}\right)^\alpha\left({1\over q^2-m_{D^*}^2
%}\right)\left[-g_{\alpha\beta}~+~{q_\alpha q_\beta\over m_{D^*}^2}\right] 
%\nonumber\\
&=&g_{\pi DD^*}~\left[ p_{1\alpha}\left({1\over q^2-m_{D^*}^2
}\right)\left(g^{\alpha\beta}~-~{q^\alpha q^\beta\over m_{D^*}^2}\right)~
+~{q^\beta\over 2m_{D^*}^2}\right]~,
\label{AI}
\eeqa
where $q=p_1-p_4$.
It is worth noting that the term proporticional to $q_\beta$ corresponds to a 
contact interaction. In the case of Eq.~(\ref{jddsu}), we get
\beqa
iA_\alpha^{(II)}~i\Delta^{\alpha\beta} &=&i g_{\pi DD^*}~ p_{1\alpha}
~i\Delta^{\alpha\beta}
\nonumber\\
&=&g_{\pi DD^*}~p_{1\alpha}\left({1\over q^2-m_{D^*}^2
}\right)\left(g^{\alpha\beta}~-~{q^\alpha q^\beta\over m_{D^*}^2}\right)~.
\label{AII}
\eeqa

The scattering amplitude is proportional to $\Gamma_\alpha A^\alpha$ and
hence, the equivalence between these two calculations requires $\Gamma^\beta
q_\beta=0$. However, using Eq.~(\ref{veve}) we find
\beq
\Gamma^\beta q_\beta=g_{\psi D^*D^*}~\epsilon_2^\beta~\epsilon_{3\beta}^*
~\left(m_\psi^2-m_{D^*}^2\right)~,
\eeq
which vanishes only in the case of exact $SU(4)$, but is different from 
zero in the case of realistic masses. The full amplitude for the
process in Fig.~1 is given by
\beqa
{\cal M}^{(II)} &=&g_{\psi D^*D^*}~g_{\pi DD^*}~\epsilon_2^\mu\epsilon_3^{*
\nu}~{1\over (p_1-p_4)^2-m_{D^*}^2}~\left[g_{\mu\nu}\left((p_2+p_3)_\alpha~
+~\left({m_\psi^2\over m_{D^*}^2}-1\right)(p_3-p_2)_\alpha\right)\right.~
\nonumber\\
&-&\left.2p_{3\mu}~
g_{\alpha\nu}~-~2p_{2\nu}~g_{\alpha\mu}\right]~p_1^\alpha~,
\label{MII}
\eeqa
and 
\beq
{\cal M}^{(I)}= {\cal M}^{(II)} ~+~{1\over2}g_{\psi D^*D^*}~g_{\pi DD^*}~
\epsilon_2^\mu\epsilon_3^{*\nu}~g_{\mu\nu}~\left({m_\psi^2\over m_{D^*}^2}
-1\right)~.
\label{MI}
\eeq

The difference between ${\cal M}^{(I)}$ and ${\cal M}^{(II)}$ is due to 
terms proportional to $\partial_\mu D^{*\mu}$ and $\partial_\mu\bar{D^{*\mu}}$
in Eq.~(\ref{rel}). Indeed, the interaction Lagrangian (\ref{jdsds})
gives rise to the following equation of motion for the $D^*$ meson:
\beqa
&-&\partial_\rho~(\partial^\rho D^{*\lambda}-\partial^\lambda  D^{*\rho})~-~
m_{D^*}^2~D^{*\lambda}~=~i~g_{\psi D^*D^*}~\left[(\partial_\rho~\psi^\rho)~
D^{*\lambda}~-~\psi^\lambda~(\partial_\rho~D^{*\rho})~\right.
\nonumber\\
&+&\left.2~\psi^\rho~
(\partial_\rho~D^{*\lambda})~
-~D^{*\rho}~(\partial_\rho~\psi^{\lambda})~-~\psi^\rho~
(\partial^\lambda~D^*_\rho)~+~D^*_\rho~(\partial^\lambda~\psi^\rho)\right]~.
\label{eqmo}
\eeqa
The equation of motion for the mesons $\bar{D^*}$ and $J/\psi$ are
totally analogous. Applying $\partial_\lambda$ into Eq.~(\ref{eqmo}), we get
\beq
-m_{D^*}^2~\partial_\lambda D^{*\lambda}~=~i~g_{\psi D^*D^*}~\left[-\psi_\rho
\partial_\lambda (\partial^\lambda ~D^{*\rho}-\partial^\rho~D^{*\lambda})~+~
\partial_\lambda (\partial^\lambda ~\psi^\rho-\partial^\rho~\psi^\lambda)~
D^*_\rho\right]~.
\eeq

Thus, using the equations of motion for $\bar{D^*}$ and $J/\psi$, and
neglecting terms proportional to $g_{\psi D^*D^*}^2$ we have
\beq
\partial_\lambda D^{*\lambda}~=~i~g_{\psi D^*D^*}~\left({m_\psi^2\over 
m_{D^*}^2}-1\right)\psi^\lambda~D^*_\lambda~.
\label{div}
\eeq
Using the above result in Eq.~(\ref{rel}) we get, to order 
 $g_{\psi D^*D^*}^2$:
\beq
{\cal L}_{\pi DD^*}^{(I)}={\cal L}_{\pi DD^*}^{(II)}-{1\over2}g_{\pi DD^*}~
g_{\psi D^*D^*}~\left({m_\psi^2\over m_{D^*}^2}-1\right)~\psi^\mu
\left(\bar D  \vec \tau D^{*}_\mu-\bar{D^{*}_\mu} \vec \tau D\right) \cdot  
\vec \pi ~.
\label{rel2}
\eeq

The second term in this equation is precisely the
effective four-leg interaction that gives rise to the difference between
${\cal M}^{(I)}$ and ${\cal M}^{(II)}$ as in Eq.~(\ref{MI}). The 
importante feature of the last term in Eq.~(\ref{rel2}) is that it breaks
$SU(2)\times SU(2)$ chiral symmetry when $SU(4)$.
As demonstrated long ago by
Weinberg \cite{wein}, the construction of chiral non-linear
Lagrangians in the $SU(2)$
sector requires necessarily gradient couplings for the pion.
In the present problem, this means that ${\cal L}_{\pi DD^*}^{(II)}$
is chiral symmetric  whereas ${\cal L}_{\pi DD^*}^{(I)}$ is not.
Consistently the former 
yields the amplitude ${\cal M}^{(II)}$, Eq.~(\ref{MII}), which is linear
in the pion momentum and would vanish if it were soft. The second term in 
Eq.~(\ref{rel2}), on the other hand, breaks chiral symmetry and gives rise
to the large amplitude ${\cal M}^{(I)}$. 

We can evaluate  numerically the difference between ${\cal M}^{(I)}$ and
${\cal M}^{(II)}$ by calculating the differential cross section. After 
including isospin factors, the differential cross section is given by
\beq
{d\sigma\over dt}={1\over 96\pi s{\bf p}_{i,cm}^2}~\sum_{spin}|{\cal M}|^2~,
\eeq
where ${\bf p}_{i,cm}$ is the three-momentum of $p_1$ (or $p_2$) in the center
of mass frame:
\beq
{\bf p}_{i,cm}^2={s^2+m_1^4+m_2^4-2sm_1^2-2sm_2^2-2m_1^2m_2^2\over4s}~,
\eeq
with $s=(p_1+p_2)^2$ and $t=(p_1-p_3)^2$.

In order to compare our results with the previous ones, we use
the same coupling constants as in refs.~\cite{mamu98,osl,linko}, namely
$g_{\pi DD^*}=8.8$ and $g_{\psi D^*D^*}=7.64$, although in a recent paper
we find a rather smaller value: $g_{\pi DD^*}=4.0\pm0.3$ 
\cite{nos}\footnote{The 
definition of $g_{\pi DD^*}$ here differs from the one used in ref.~\cite{nos}
by a factor $1/\sqrt{2}$.}.
In Fig.~2 we show the cross section for the $\pi+J/\psi\rightarrow D+D^*$
process obtained from Fig.~1 with  the Lagrangian, 
${\cal L}_{\pi DD^*}^{(II)}$,
in  Eq.~(\ref{jddsu}) (dashed line), and with the Lagrangian, ${\cal L}_{\pi 
DD^*}^{(I)}$, in Eq.~(\ref{jddko}) (dash-dotted line). As expected,
the cross section obtained with ${\cal L}_{\pi DD^*}^{(I)}$, which breaks 
chiral $SU(2)\times SU(2)$, is much bigger
than that obtained with the chiral Lagrangian ${\cal L}_{\pi 
DD^*}^{(II)}$. The difference between both results is even more
important near threshold where the cross section obtained with ${\cal L}_{\pi 
DD^*}^{(I)}$ grows very rapidly. For instance, at $\sqrt{s}=4 \GeV$ the chiral
Lagrangian gives $\sigma\sim3.5$ mb, while ${\cal L}_{\pi DD^*}^{(I)}$
gives  $\sigma\sim11.5$ mb. In Fig.~2 we also show, for completeness,
the cross section for the $\pi+J/\psi\rightarrow D+D^*$
process obtained with a $D$ meson exchange (solid line). In this case the 
result is the same with both Lagrangians since the external vector mesons
are free and the conditions $\partial_\mu 
D^{* \mu}=\partial_\mu \bar{D^{* \mu}}=0$ hold.

%\section{Conclusions}

In conclusion, we have shown that the two alternative forms of the $\pi DD^*$
interaction Lagrangian found in the literature are not equivalent for
processes involving a virtual $D^*$ meson exchange. In particular
the form which breaks chiral $SU(2)\times SU(2)$ symmetry gives a bigger
cross section, this effect being even more important near threshold.
Since $SU(4)$ symmetry is strongly broken in nature, we believe that
the interaction Lagrangians involving the pion should be constructed respecting
chiral symmetry in the $SU(2)$
sector, which requires necessarily gradient couplings for the pion.

\vspace{1.0cm}

\underline{Acknowledgements}: We would like to thank Che Ming Ko for 
useful discussions. This work has been supported by CNPq and  
FAPESP under contract number 1999/12987-5.

%\newpage
   
%\section*{Appendix}
%\appeqn

\begin{figure} \label{fig1}
\begin{center}
%\vskip -1cm
\epsfysize=6.0cm
\epsffile{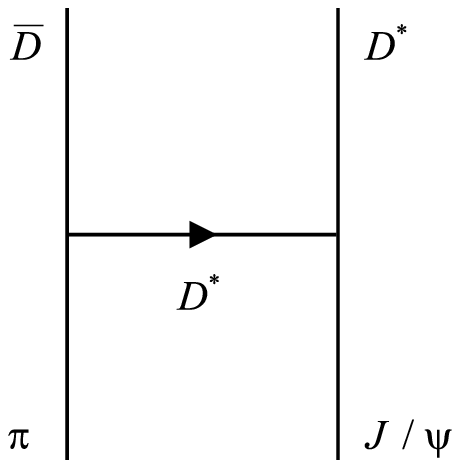}
\caption{Diagram for the process $\pi+J/\psi\rightarrow D+D^*$ with 
the $D^*$ meson exchange.}
\end{center}
\end{figure}

\begin{figure} \label{fig2}
\begin{center}
%\vskip -1cm
\epsfysize=8.0cm
\epsffile{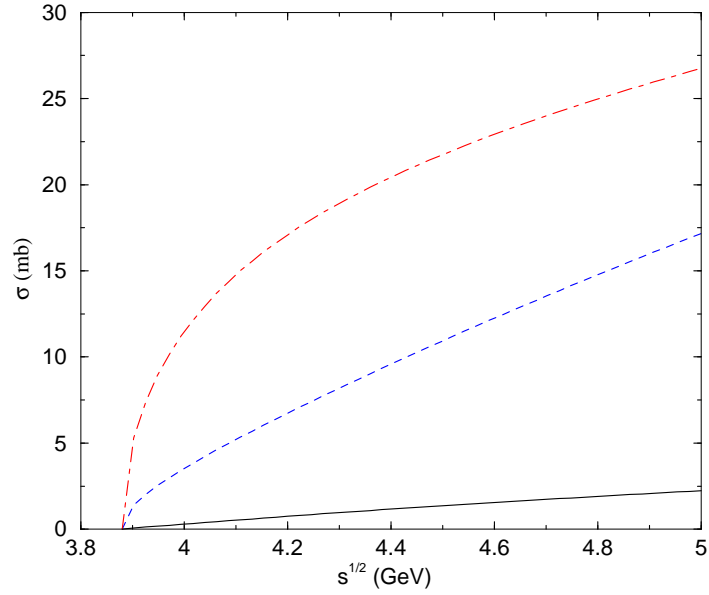}
\caption{Cross sections for the process $\pi+J/\psi\rightarrow D+D^*$ with:
$D$ meson exchange (solid line), $D^*$ meson exchange with the Lagrangian 
Eq.~(\protect\ref{jddsu}) (dashed line), and with the Lagrangian 
Eq.~(\protect\ref{jddko}) (dash-dotted line).}
\end{center}
\end{figure}

\end{document}